\documentclass{article}

\usepackage{graphicx} 
\usepackage{subfigure} 

\usepackage{natbib}

\usepackage{algorithm}
\usepackage{algorithmic}

\usepackage{hyperref}

\usepackage{times}
\usepackage{helvet}
\usepackage{courier}
\usepackage{color}
\usepackage{colortbl}
\usepackage{algorithm}
\usepackage{algorithmic}
\usepackage{amsmath}
\usepackage{amssymb}
\usepackage{latexsym} 
\usepackage{multirow}
\usepackage{verbatim}
\usepackage{amsthm}

\newcommand{\II}{\mathcal{I}}

\newcommand{\EV}{\mathbb{E}}

\newcommand{\abs}[1]{\left|#1\right|}
\newcommand{\Qed}{$\blacksquare$}
\newtheorem{definition}{Definition}

\newtheorem{theorem}{Theorem}

\newcommand{\defword}[1]{\textbf{\boldmath{#1}}}

\def\etal{\textit{et al.}}
\def\ie{\textit{i.e.}}

\newtheorem{df}{Definition}
\newtheorem*{definition4}{Definition 4}

\newtheorem*{theorem2}{Theorem 2}
\newtheorem*{theorem3}{Theorem 3}

\newtheorem*{lemmaA}{Lemma A}

\newcommand{\bt}{\begin{theorem}\em}
\newcommand{\et}{\end{theorem}}

\DeclareMathOperator*{\argmax}{arg\,max}
\DeclareMathOperator*{\argmin}{arg\,min}

\def\etal{\textit{et al.}}
\def\ie{\textit{i.e.}}

\newcommand{\bea}{\begin{eqnarray}}
\newcommand{\eea}{\end{eqnarray}}

\newcommand{\bdf}{\begin{df}\em}
\newcommand{\edf}{\end{df}}

\newcommand{\ben}{\begin{enumerate}}
\newcommand{\een}{\end{enumerate}}

\definecolor{grey}{rgb}{0.4,0.4,0.4}
\definecolor{loselots}{rgb}{1,0.4,0.4}
\definecolor{losesome}{rgb}{1,0.6,0.6}
\definecolor{losebit}{rgb}{1,0.8,0.8}
\definecolor{tie}{rgb}{1,1,1}
\definecolor{winbit}{rgb}{0.8,1,0.8}
\definecolor{winsome}{rgb}{0.6,1,0.6}
\definecolor{winlots}{rgb}{0.4,1,0.4}

\title{No-Regret Learning in Extensive-Form Games with Imperfect Recall}

\author{\hspace{-1.5cm}Marc Lanctot$^1$, Richard Gibson$^1$, Neil Burch$^1$, 
Martin Zinkevich$^2$, and Michael Bowling$^1$\\
\hspace{-1.5cm}$^1$ Department of Computing Science, University of Alberta\\
\hspace{-1.5cm}$^2$ Yahoo! Reseach\\}

\date{}

\newif\iftechreport
\techreporttrue

\begin{document} 


\maketitle


\begin{abstract} 
Counterfactual Regret Minimization (CFR) is an efficient no-regret learning algorithm for decision problems modeled as extensive games. 
CFR's regret bounds depend on the requirement of perfect recall: players always remember information that was revealed to them and the order in which it was revealed. 
In games without perfect recall, however, CFR's guarantees do not apply.
In this paper, we present the first regret bound for CFR when applied
to a general class of games with imperfect recall. 
In addition, we show that CFR applied to any abstraction
belonging to our general class results in a regret bound not just for the abstract game, but for the full game as well.
We verify our theory and show how imperfect recall can be used to trade a small increase in regret for a significant reduction in memory in three domains: die-roll poker, phantom tic-tac-toe, and Bluff. 
\end{abstract}

\section{Introduction}

Many real-world problems can be modeled as a repeated decision-making task. 
For problems involving multiple agents, one can model the repeated task as a normal-form game. 
When the task incorporates sequential decisions involving imperfect information or stochastic events, an extensive game is a useful alternative. 
In such decision problems, a typical goal is to minimize regret: the
amount of utility lost by playing a past sequence of strategies,
versus playing the best, stationary strategy in hindsight. 

In this paper, we consider the problem of minimizing regret in an extensive game. 
A common approach to achieving low regret in extensive games is the Counterfactual Regret Minimization (CFR)~\citep{CFR} algorithm. 
CFR uses a regret minimizer at every decision point with an
alternative notion of regret, which provably minimizes regret in the entire extensive game. 
However, convergence is limited to games exhibiting perfect recall: players never forget information that was revealed to them, nor the order in the which the information was revealed. 
For games with imperfect recall, CFR's original analysis provides no general guarantees. 

Imperfect recall brings about a number of complications. 
In games with perfect recall, every mixed strategy (probability distribution over pure strategies) has a utility-equivalent behavioral strategy (probability distribution over actions at each decision point)~\citep{Kuhn53}. 
While certain lossless imperfect recall games share this property~\citep{KanekoKline95}, it is not true for imperfect recall games in general~\citep{Piccione96}. 
In addition, the decision problem of determining if a player can assure themself a certain payoff in an imperfect recall game is NP-complete~\citep{Koller92the}. 
Two-player zero-sum games can be solved by constructing an appropriate linear program~\citep{SequenceFormLPs} or minimizing regret~\citep{CFR}, provided the game has perfect recall. 
Without perfect recall, however, the problem becomes exponential in the worst case~\citep{SequenceFormLPs}. 

On the other hand, imperfect recall extensive games are more versatile than perfect recall games for modelling large real-world problems. 
While perfect recall requires all past information to be remembered, imperfect recall allows irrelevant information to be forgotten so that the size of the game is smaller. 
As CFR's memory requirements are linear in the size of the game, more games become feasible through imperfect recall. 
Despite the complications above, CFR has empirically been shown to work well when applied to imperfect recall abstractions of Texas Hold'em poker~\citep{Waugh09}, but there is currently no theory to suggest why this is so. 

This paper presents theoretical groundings for applying CFR to games exhibiting imperfect recall. 
We define a general class of imperfect recall games and provide a
bound on CFR's regret in such games. 
For a subset of this class, CFR minimizes average regret in the extensive game. 
Moreover, our results also provide regret guarantees when applying
CFR to an abstract game, provided the abstract game belongs to our general class. 
We test our theory in three different domains: die-roll poker, phantom tic-tac-toe, and Bluff. 
To the best of our knowledge, this work demonstrates the first theoretically-grounded, practical use of imperfect recall in extensive games.

\section{Background \label{sec:background}}

An extensive-form game $\Gamma$ with imperfect information~\citep{OsbRub94} is a tuple $\langle N, A, H, Z, P, \sigma_c,$ $u, \II \rangle$, where
$N$ is a finite set of {\bf players}. 
$A$ is a finite set of {\bf actions}. 
$H$ is a finite set of {\bf histories}: a subset of the set of sequences of elements in $A$. 
      A {\bf prefix} of a history $h' \in H$ is a history $h \in H$ where $h'$ begins with the sequence $h$; we denote prefix histories by $h \sqsubseteq h'$. 
      For every $h \in H$, define $A(h) = \{ a : a \in A, ha \in H \}$, 
      the set of valid actions at history $h$; $P(h) \in N \cup \{
      c \}$ is the player to act at the history $h$, or {\it chance} if
      $P(h) = c$; and $H_i = \{h \mid h \in H, P(h) = i\}$. 
$Z \subseteq H$ is the set of {\bf terminal histories}. A terminal history $z \in Z$ is a history where
      there does not exist any history $h \in H$, $h \neq z$ such that 
      $z \sqsubseteq h$. 
      The {\bf utility function} $u_i: Z
      \rightarrow \mathbb{R}$ gives the utility to player $i \in N$,
      for each terminal history. If $|N|=2$ and for all $z \in Z$, $\sum_{i\in N} u_i(z) = 0$, we say the game is {\bf zero-sum}.

For each player $i \in N$, $\II_i$ is a partition of $H_i$ with  
      the property that $A(h) = A(h')$ whenever $h$
      and $h'$ are in the same member of the partition. We call
      $\II_i$ the {\bf information partition} of player $i$, and a set $I \in \II_i$
      is an {\bf information set} of player $i$.  A player, when taking actions, cannot distinguish between two histories in the same information set.
      For $I \in \II_i$ we denote $A(I)$ as the set $A(h)$ for 
      any $h \in I$. Define $I(h)$ to be the information set containing $h$. 
In this paper, we restrict ourselves to games where players cannot reach the same information set twice in a single game. 
Thus, we assume that for all $i \in N$ and $h,h' \in H_i$,
\begin{equation}
\label{eq:noloops}
h \sqsubseteq h', h \neq h' \Rightarrow I(h) \neq I(h'). 
\end{equation}

Finally, $\sigma_c$ is the fixed ``strategy'' of the special player {\it
  chance}.  $\sigma_c(h,a)$ gives the probability that
chance event $a$ occurs at $h$.  For all $h \in H_c$, $\sum_{a
  \in A(h)}\sigma_c(h,a)=1$ and the decisions at any $h$ are
independent of the decision at any other $h' \ne h$. 

Given a history $h$, define $X_i(h)$ to be the sequence of information set, action pairs such that 
$(I,a) \in X_i(h)$ if $I \in \II_i$ and there exists $h' \sqsubseteq h$ such that $h' \in I$ and 
$h'a \sqsubseteq h$. The order of the pairs in $X_i(h)$ is the order in which they occur in $h$. 
Define $X(h)$ to be the sequence of information set, action pairs belonging to all players
in the order in which they occur in $h$, and $X_{-i}(h)$ similarly, by 
removing player $i$'s information set, action pairs from $X(h)$. 
Also, define $X(h,h')$ to be the sequence of information set, action pairs belonging to all players that start at $h$ and end at $h'$ when $h \sqsubseteq h'$; if $h \not \sqsubseteq h'$, $X(h,h')$ is defined to be the empty sequence.
Finally, $X_i(h,h')$ and $X_{-i}(h,h')$ are similarly defined.

\begin{definition}
An extensive game has {\bf perfect recall} if for every player $i \in N$, for every information set $I \in \II_i$, 
for any $h, h' \in I: X_i(h) = X_i(h')$.  Otherwise, the game has {\bf imperfect recall}.
\end{definition}
Intuitively, with perfect recall every player has an infallible memory: they cannot ``forget'' anything during a play of the
game that they once knew. Hence, what a player knows at $I$ is a composition of what the player has 
discovered in the past up to this point and the precise order in which information was discovered. 
Note that every perfect recall game satisfies equation \eqref{eq:noloops}, but not every imperfect recall game does. 

A \defword{(behavioral) strategy} $\sigma_i$ for player $i$ is a function such that for each history $h \in H_i$, $\sigma_i(h)$ is a probability distribution over $A(h)$.  Furthermore, it is required that $\sigma_i(h) = \sigma_i(h')$ for all $h,h' \in I$, and we denote that as $\sigma_i(I)$. 
The set of all such strategies for player $i$ is denoted by $\Sigma_i$. 
A {\bf strategy profile} $\sigma \in \Sigma$ is a collection of strategies, one for each player, {\it i.e.} in a two-player game $\sigma = (\sigma_1, \sigma_2)$.
By notational convention, $\sigma_{-i}$ refers to the set of strategies including every strategy in $\sigma$ except player $i$'s strategy. 

For any $\sigma \in \Sigma$, $i \in N \cup \{ c \}$, and $h \in H$, define $\pi_i^\sigma(h) = \prod_{h'a\sqsubseteq h, P(h') = i}\sigma_i(h',a)$ to be the probability that player $i$ plays to reach history $h$ under $\sigma$.
We can then define $\pi^\sigma(h) = \prod_{i \in N \cup \{ c\}}\pi_{i}^\sigma(h)$ to be the probability that history $h$ is reached under $\sigma$.
Let $\pi_{-i}^\sigma(h)$ be the product of all players' contribution (including chance) except that of player $i$.
Furthermore, let $\pi_i^\sigma(h,h')$ be the probability of player $i$ playing to reach history $h'$ after $h$, given $h$ has occurred.
Let $\pi^\sigma(h,h')$ and $\pi_{-i}^\sigma(h,h')$ be defined similarly.
Finally, we can define the expected utility of a strategy profile $\sigma$ for player $i$ to be
\[u_i(\sigma) = \EV_{z \in Z}[u_i(z)] = \sum_{z \in Z} u_i(z) \pi^\sigma(z). \]

We will say that a game $\Gamma' = \langle N, A', H, Z, P, \sigma_c, u,
\II' \rangle$ is an \defword{abstraction}, or an \defword{abstract game}, of $\Gamma = \langle N, A, H, Z, P, \sigma_c, u,
\II \rangle$ if for all $i \in N$ and $h,k \in H_i$, $A'(h) \subseteq A(h)$
and $I(h)=I(k)$ implies $I'(h)=I'(k)$. 
In this paper, we only consider abstractions where $A = A'$. 
A typical use of abstraction is to reduce the size of the game by ensuring that $|\II'| < |\II|$.  


\section{Example: Die-Roll Poker}

We now introduce a game that we will use as a running example throughout the paper. 

\textbf{Die-roll poker (DRP)} is a simplified two-player poker game that uses dice rather than cards.
To begin, each player antes one chip to the \textbf{pot}. 
There are two betting rounds, where at the beginning of each round, players roll a private six-sided die.
The game has imperfect information due to the players not seeing the result of the opponent's die rolls.
During a betting round, a player may \textbf{fold} (forfeit the game), \textbf{call} (match the current bet), or \textbf{raise} (increase the current bet) by a fixed number of chips, with a maximum of two raises per round. 
In the first round, raises are worth two chips, whereas in the second round, raises are worth four chips. 
If both players have not folded by the end of the second round, a \defword{showdown} occurs where the player with the largest sum of their two dice wins all of the chips in the pot.

DRP is naturally a game with perfect recall; players remember the exact sequence of bets made and the exact outcome of each die roll from both rounds. 
However, consider an imperfect recall version of DRP, \textbf{DRP-IR}, where at the beginning of the second round, both players ``forget'' their first die roll and only know the sum of their two dice.
In other words, DRP-IR is an abstraction of DRP where any two histories are in the same abstract information set if and only if the sum of the player's private dice is the same and the sequence of betting is the same. 
DRP-IR has imperfect recall since histories that were distinguishable in the first round (for example, a roll of 1 and a roll of 4) are no longer distinguishable in the second round (for example, a roll of 1 followed by a roll of 5, and a roll of 4 followed by a roll of 2).

\section{Counterfactual Regret Minimization \label{sec:cfr}}

Given a sequence of strategy profiles $\sigma^1, \sigma^2, ..., \sigma^T$, the \defword{(external) regret} for player $i$,
\[ R_i^T = \max_{\sigma' \in \Sigma_i} \sum_{t=1}^T \left( u_i(\sigma', \sigma_{-i}^t) - u_i(\sigma_i^t, \sigma_{-i}^t) \right), \]
is the amount of utility player $i$ could have gained had she played the best single strategy in hindsight for all time steps $t \in \{1, 2, ..., T\}$. 
An algorithm \defword{minimizes regret}, or is a \defword{no-regret algorithm}, for player $i$ if the average positive regret approaches zero; \ie,~$\lim_{T \rightarrow \infty} R_i^{T,+} / T = 0$, where $x^+ = \max\{x, 0\}$. 
Having no regret is a desirable property.
For example, it is well known that in a zero-sum game, if both players' average regret is bounded above by $\epsilon$, then the average of the strategy profiles generated is a $2\epsilon$-Nash equilibrium. 

\defword{Counterfactual Regret Minimization (CFR)} is an iterative no-regret learning algorithm for extensive-form games having perfect recall. 
On each iteration $t$, CFR recursively traverses the entire game tree, computing the expected utility for player $i$ at each information set $I \in \mathcal{I}_i$ under the current profile $\sigma^t$, assuming player $i$ plays to reach $I$. 
This expectation is the \defword{counterfactual value} for player $i$,
\[ v_i(\sigma, I) = \sum_{z \in Z_I} u_i(z)\pi_{-i}^{\sigma}(z[I])\pi^{\sigma}(z[I],z), \]
\noindent where $Z_I$ is the set of terminal histories passing through $I$ and $z[I]$ is the prefix of $z$ contained in $I$ ($z[I]$ is unique by equation \eqref{eq:noloops}).  
For each action $a \in A(I)$, these values determine the \defword{counterfactual regret} at iteration $t$, $r_i^t(I,a) = v_i(\sigma^t_{I \rightarrow a}, I) - v_i(\sigma^t, I)$, where $\sigma_{I \rightarrow a}$ is the profile $\sigma$ except at $I$, action $a$ is always taken.  
The regret $r_i^t(I,a)$ measures how much player $i$ would rather play action $a$ at $I$ than play $\sigma^t$.  Finally, $\sigma^t$ is updated by applying regret matching~\citep{Hart00,CFR} to the \defword{immediate counterfactual regrets}, $R_i^T(I,a) = \sum_{t=1}^T r_i^t(I,a)$, according to
\[ \sigma^{T+1}(I,a) = \frac{R_i^{T,+}(I,a)}{\sum_{b \in A(I)} R_i^{T,+}(I,b)}, \]
 with actions chosen uniformly at random when the denominator is zero. 
Regret matching is a no-regret learner that minimizes the per-information set immediate counterfactual regret,
\begin{equation}
\label{eq:blackwell}
\max_{a \in A(I)} \frac{R_i^T(I,a)}{T} \leq \frac{\Delta_{i} \sqrt{|A(I)|}}{\sqrt{T}},
\end{equation}
where $\Delta_{i} = \max_{z, z' \in Z} u_i(z) - u_i(z')$. 
In games having perfect recall, minimizing the immediate counterfactual regrets at every information set in turn minimizes average regret, $R_i^T / T$. 
This is because perfect recall implies that the regret is bounded by the sum of the positive parts of the immediate counterfactual regrets \citep{CFR},
\begin{equation}
\label{eq:CFR}
R_i^T \leq \sum_{I \in \II_i} \max_{a \in A(I)} R_i^{T,+}(I,a),
\end{equation}
and thus
\begin{equation}
\label{eq:RegretBound}
\frac{R_i^T}{T} \leq \frac{\Delta_{i} |\II_i| \sqrt{|A_i|}}{\sqrt{T}},
\end{equation}
where $|A_i| = \max_{I \in \II_i} \abs{A(I)}$. 
CFR must store the immediate counterfactual regret for each information set, action pair, and thus CFR's memory requirements are $O(|\mathcal{I}_i||A_i|)$. 

While equation \eqref{eq:blackwell} still holds in imperfect recall games, equation~\eqref{eq:CFR} and consequently equation \eqref{eq:RegretBound} are not guaranteed to hold. 
An example game where CFR would exhibit high regret is provided in Section~\ref{sec:discussion}. 
Consequently, the regret for playing according to the CFR algorithm is unknown in general for imperfect recall games. 
However, the advantage of applying CFR to DRP-IR, for example, is that this imperfect recall game contains fewer information sets than the full game, and thus less memory is required by CFR. 
Although DRP is a toy example and is small enough to run CFR on the full game, this example is useful for understanding the concepts in the rest of this paper. 

\section{CFR with Imperfect Recall}
\label{sec:theory}

In this section, we investigate the application of CFR to games with imperfect recall. 
We begin by showing that CFR minimizes regret for a class of games that we call ``well-formed games.'' 
We then present a bound on the average regret for a more general class of imperfect recall games that we call ``skew well-formed games.'' 

\subsection{Well-formed Games}

For games $\Gamma = \langle N, A, H, Z, P, \sigma_c, u, \II \rangle$ and $\breve{\Gamma} = \langle N, A, H, Z, P, \sigma_c, u, \breve{\II} \rangle$, we say that $\breve{\Gamma}$ is a \defword{perfect recall refinement of $\Gamma$} if $\breve{\Gamma}$ has perfect recall and $\Gamma$ is an abstraction of $\breve{\Gamma}$.  So, the information available to players in $\breve{\Gamma}$ is never forgotten, and is at least as informative as the information available to them in $\Gamma$.
For example, DRP is a perfect recall refinement of DRP-IR. 
Every game has at least one perfect recall refinement by simply making $\breve{\Gamma}$ a perfect information game ($\breve{I} = \{h\}$ for all $\breve{I} \in \breve{\II}_i)$. 
Furthermore, a perfect recall game is a perfect recall refinement of itself. 
For $I \in \II_i$, we define 
\[ \breve{\mathcal{P}}(I) = \{ \breve{I} \mid \breve{I} \in \breve{\II}_i, \breve{I} \subseteq I \} \] 
to be the set of all information sets in $\breve{\II}_i$ that are subsets of $I$. 
Note that our notion of refinement is similar to the one described by Kaneko \& Kline (\citeyear{KanekoKline95}). 
Our definition differs in that we consider any possible refinement, whereas Kaneko \& Kline consider only the coarsest such refinement. 

\begin{definition}
\label{def:wellformed}
For a game $\Gamma$ and a perfect recall refinement $\breve{\Gamma}$, we say that $\Gamma$ is a \defword{well-formed game with respect to $\breve{\Gamma}$} if for all $i \in N$, $I \in \mathcal{I}_i$, $\breve{I}, \breve{I}' \in \breve{\mathcal{P}}(I)$, there exists a bijection $\phi: Z_{\breve{I}} \rightarrow Z_{\breve{I}'}$ and constants $k_{\breve{I},\breve{I}'}, \ell_{\breve{I},\breve{I}'} \in [0, \infty)$ such that for all $z \in Z_{\breve{I}}$:
\begin{enumerate}
\item[\emph{(i)}] $u_i(z) = k_{\breve{I},\breve{I}'}u_i(\phi(z))$,
\item[\emph{(ii)}] $\pi_c(z) = \ell_{\breve{I},\breve{I}'} \pi_c(\phi(z))$,
\item[\emph{(iii)}] In $\Gamma$, $X_{-i}(z) = X_{-i}(\phi(z))$, and 
\item[\emph{(iv)}] In $\Gamma$, $X_i(z[\breve{I}], z) = X_i(\phi(z)[\breve{I}'], \phi(z))$.
\end{enumerate}
We say that $\Gamma$ is a \defword{well-formed game} if it is well-formed with respect to some perfect recall refinement.
\end{definition}
Recall that $Z_I$ is the set of terminal histories containing a prefix in the information set $I$, and that $z[I]$ is that prefix. 
Intuitively, a game is well-formed if for each information set $I \in \II_i$, the structures around each $\breve{I},\breve{I}' \in \breve{\mathcal{P}}(I)$ of some perfect recall refinement are isomorphic across four conditions.
Conditions (i) and (ii) state that the corresponding utilities and chance frequencies at each terminal history are proportional.
Condition (iii) asserts that the opponents can never distinguish the corresponding histories at any point in $\Gamma$.
Finally, condition (iv) states that player $i$ cannot distinguish between corresponding histories from $\breve{I}$ and $\breve{I}'$ until the end of the game.

Consider again DRP as a perfect recall refinement of DRP-IR. 
In DRP, the available actions are independent of dice outcomes, and the final utilities are only dependent on the final sum of the players' dice. 
Therefore, in DRP the utilities are equivalent between, for example, the terminal histories where player $i$ rolled a 1 followed by a 5, and the terminal histories where player $i$ rolled a 4 followed by a 2 (condition (i)).
In addition, the chance probabilities of reaching each terminal history are equal (condition (ii)).
Furthermore, the opponents can never distinguish between two isomorphic histories since player $i$'s rolls are private (condition (iii)).
Finally, in DRP-IR, player $i$ never remembers the outcome of the first roll from the second round on (condition (iv)).
Thus, DRP-IR is well-formed with respect to DRP, with constants $k_{\breve{I},\breve{I}'} = \ell_{\breve{I},\breve{I}'} = 1$.

Any perfect recall game is well-formed with respect to itself since $\breve{\mathcal{P}}(I) = \{I\}$, $\phi$ equal to the identity bijection, and $k_{\breve{I},\breve{I}'} = \ell_{\breve{I},\breve{I}'} = 1$ satisfies Definition \ref{def:wellformed}. 
However, many imperfect recall games are also well-formed, with DRP-IR being one example. 
An additional example is presented in Section \ref{sec:eval}. 

We now show that CFR can be applied to any well-formed game to minimize average regret. 
A sketch of the proof is described below, while a full proof is provided as supplementary material. 
\begin{theorem}
\label{thm:safe}
If $\Gamma$ is well-formed with respect to $\breve{\Gamma}$, then the average regret in $\breve{\Gamma}$ for player $i$ of choosing strategies according to CFR in $\Gamma$ is bounded by
\[ \frac{\breve{R}_i^T}{T} \leq \frac{\Delta_{i} K \sqrt{|A_i|}}{\sqrt{T}}, \]
where $K = \sum_{I \in \mathcal{I}_i} \max_{\breve{I},\breve{I}' \in \breve{\mathcal{P}}(I)} k_{\breve{I},\breve{I}'} \ell_{\breve{I},\breve{I}'}$.
\end{theorem}
\textbf{Proof sketch.} 
One can show that conditions (i) to (iv) of Definition \ref{def:wellformed} imply that the positive regrets are proportional between any two information sets in $\breve{\Gamma}$ that are merged in the well-formed game, $\Gamma$. 
In other words, for all $I \in \mathcal{I}_i$, $\breve{I},\breve{I}' \in \breve{\mathcal{P}}(I)$, and $a \in A(I)$, 
\[ R_i^{T,+}(\breve{I},a) = k_{\breve{I},\breve{I}'} \ell_{\breve{I},\breve{I}'} R_i^{T,+}(\breve{I}',a). \]
Since regrets between $\Gamma$ and $\breve{\Gamma}$ are additive, \ie,
\[ R_i^T(I,a) = \sum_{\breve{I} \in \breve{\mathcal{P}}(I)} R_i^T(\breve{I},a) \text{ for all } I \in \II_i, \]
the proportionality implies that minimizing regret at each $I \in \mathcal{I}_i$ minimizes regret at each $\breve{I} \in \breve{\II}_i$. 
Because $\breve{\Gamma}$ has perfect recall, applying equation \eqref{eq:CFR} gives the result. \Qed

Since the strategy space is more expressive in $\breve{\Gamma}$ than in $\Gamma$ ($\Sigma \subseteq \breve{\Sigma}$), $R_i^T \leq \breve{R}_i^T$ and thus it immediately follows that the average regret in $\Gamma$ is minimized. 
In the case when $\Gamma$ has perfect recall, because $\Gamma$ is well-formed with respect to itself, Theorem 1 with $K = |\mathcal{I}_i|$ is a direct generalization of the original CFR bound in equation \eqref{eq:RegretBound}. 
Theorem \ref{thm:safe} not only guarantees regret minimization for perfect recall games, but also for well-formed imperfect recall games. 

\subsection{Skew Well-formed Games}

We now present a generalization of well-formed games to which a regret bound can still be derived. 
\begin{definition}
\label{def:skewwellformed}
For a game $\Gamma$ and a perfect recall refinement $\breve{\Gamma}$, we say that $\Gamma$ is a \defword{skew well-formed game with respect to $\breve{\Gamma}$} if for all $i \in N$, $I \in \mathcal{I}_i$, $\breve{I}, \breve{I}' \in \breve{\mathcal{P}}(I)$, there exists a bijection $\phi: Z_{\breve{I}} \rightarrow Z_{\breve{I}'}$ and constants $k_{\breve{I},\breve{I}'}, \delta_{\breve{I},\breve{I}'}, \ell_{\breve{I},\breve{I}'} \in [0, \infty)$ such that for all $z \in Z_{\breve{I}}$:
\begin{enumerate}
\item[\emph{(i)}] $\left| u_i(z) - k_{\breve{I},\breve{I}'}u_i(\phi(z)) \right| \leq \delta_{\breve{I},\breve{I}'}$,
\item[\emph{(ii)}] $\pi_c(z) = \ell_{\breve{I},\breve{I}'} \pi_c(\phi(z))$,
\item[\emph{(iii)}] In $\Gamma$, $X_{-i}(z) = X_{-i}(\phi(z))$, and 
\item[\emph{(iv)}] In $\Gamma$, $X_i(z[\breve{I}], z) = X_i(\phi(z)[\breve{I}'], \phi(z))$.
\end{enumerate}
We say that $\Gamma$ is a \defword{skew well-formed game} if it is skew well-formed with respect to some perfect recall refinement.
\end{definition}
The only difference between Definitions \ref{def:wellformed} and \ref{def:skewwellformed} is in condition (i). 
While utilities must be exactly proportional in a well-formed game, utilities in a skew well-formed game must only be proportional up to a constant $\delta_{\breve{I},\breve{I}'}$. 
Note that any well-formed game is skew well-formed by setting $\delta_{\breve{I},\breve{I}'} = 0$. 

For example, consider a new version of DRP called \defword{Skew-DRP($\delta$)} with slightly modified payouts at the end of the game. 
Whenever the game reaches a showdown, player 1 receives a bonus $\delta$ times the number of chips in the pot from player 2 if player 1's second die roll was even; otherwise, no bonus is awarded. 
The pot is then awarded to the player with the highest dice sum as usual. 
Analogously, define \defword{Skew-DRP-IR($\delta$)} to be the imperfect recall abstraction of Skew-DRP($\delta$) where in the second round, players only remember the sum of their two dice. 
Now, Skew-DRP-IR($\delta$) is not well-formed with respect to Skew-DRP($\delta$).  
To see this, note that the utilities resulting from the rolls 1,5 and the rolls 4,2 and the same sequence of betting are not exactly proportional because the second roll 5 is odd but 2 is even (utilities are off by $\delta$ times the pot size). 
However, Skew-DRP-IR($\delta$) is skew well-formed with respect to Skew-DRP($\delta$) with $\delta_{\breve{I},\breve{I}'} = \delta$ times the maximum pot size attainable from $I$. 

Unfortunately, there is no guarantee that regret will be minimized by CFR in a skew well-formed game. 
However, we can still bound regret in a predictable manner according to the degree that the utilities are skewed:
\begin{theorem}
\label{thm:skew}
If $\Gamma$ is skew well-formed with respect to $\breve{\Gamma}$, then the average regret in $\breve{\Gamma}$ for player $i$ of choosing strategies according to CFR in $\Gamma$ is bounded by
\[ \frac{\breve{R}_i^T}{T} \leq \frac{\Delta_{i} K \sqrt{|A_i|}}{\sqrt{T}} + \sum_{I \in \mathcal{I}_i} |\breve{\mathcal{P}}(I)| \delta_I, \]
where $K = \sum_{I \in \mathcal{I}_i} \max_{\breve{I},\breve{I}' \in \breve{\mathcal{P}}(I)} k_{\breve{I},\breve{I}'} \ell_{\breve{I},\breve{I}'}$ and $\delta_I = \max_{\breve{I},\breve{I}' \in \breve{\mathcal{P}}(I)} \delta_{\breve{I},\breve{I}'}\ell_{\breve{I},\breve{I}'}$.
\end{theorem}
The proof is similar to that of Theorem \ref{thm:safe}. 
Theorem \ref{thm:skew} shows that as $T$ approaches infinity, the bound on our regret approaches $\sum_{I \in \mathcal{I}_i} |\breve{\mathcal{P}}(I)| \delta_I$. 
Our experiments in Section \ref{sec:eval} demonstrate that as the skew $\delta$ grows, so does our regret in Skew-DRP($\delta$) after a fixed number of iterations. 

\textbf{Remarks.} Theorems \ref{thm:safe} and \ref{thm:skew} are, to our knowledge, the first to provide such theoretical guarantees in imperfect recall settings. 
However, these results are also relevant with regards to regret in the full game when CFR is applied to an abstraction. 
Recall that if $\Gamma$ has perfect recall, then $\Gamma$ is a perfect recall refinement of any (skew) well-formed abstract game. 
Thus, if we choose an abstraction that yields a (skew) well-formed game, then applying CFR to the abstract game achieves a bound on the average regret \emph{in the full game}, $\Gamma$. 
This is true regardless of whether the abstraction exhibits perfect recall or imperfect recall. 
Previous counterexamples show that abstraction in general provides no guarantees in the full game~\citep{AbstractionPathologies}.
In contrast, our results show that applying CFR to an abstract game leads to bounded regret in the full game, provided we restrict ourselves to (skew) well-formed abstractions. 
If such an abstract game is much smaller than the full game, a significant amount of memory is saved when running CFR. 

\section{Empirical Evaluation}
\label{sec:eval}
 
To complement our theoretical results, we apply CFR to both players simultaneously in several zero-sum imperfect recall (abstract) games, and measure the sum of the average regrets for both players in a perfect recall refinement (the full game). 
Along with the small DRP domain and its variants, we also consider the challenging domains of phantom tic-tac-toe and Bluff, which we now describe.



\textbf{Phantom tic-tac-toe.} As in regular tic-tac-toe, \defword{phantom tic-tac-toe (PTTT)} is played on a 3-by-3 board, initially empty, where the goal is to claim three squares along the same row, column, or diagonal. 
However, in PTTT, players' actions are private. 
Each turn, a player attempts to take a square of their choice. 
If they fail due to the opponent having taken that square on a previous turn, the same player keeps trying to take an alternative square until they succeed. 
Players are not informed about how many attempts the opponent made before succeeding. 
The game ends immediately if there is ever a connecting line of squares belonging to the same player. 
The winner receives a payoff of $+1$, while the losing player receives $-1$. 
In PTTT, the total number of histories $|H| \approx 10^{10}$. 

\textbf{Bluff.} Bluff, also known as Liar's Dice, Dudo, and Perudo, is a dice-bidding game. 
In our version, \defword{Bluff($D_1$,$D_2$)}, each die has six sides with faces 1 to 6. 
Each player $i$ rolls $D_i$ of these dice and looks at them without showing them to the opponent. 
Each round, players alternate by bidding on the outcome of all dice in play until one player claims that the other is bluffing (\ie, claims that the bid does not hold). 
A bid consists of a \defword{quantity} of dice and a \defword{face} value.  
A face of 6 is considered ``wild'' and counts as matching any other face. 
For example, the bid 2x5 represents the claim that there are at least two dice with a face of 5 (or 6) among both players' dice. 
To place a new bid, the player must increase either the quantity or face value of the current bid; in addition, lowering the face is allowed if the quantity is increased. 
The player calling bluff wins the round if the opponent's last bid is incorrect, and loses otherwise. 
The losing player removes one of their dice from the game and a new round begins, starting with the player who won the previous round. 
When a player has no more dice left, they have lost the game. 
A utility of $+1$ is given for a win and $-1$ for a loss. 
In this paper, we restrict ourselves to the case where $D_1 = D_2 = 2$.  
Note that since Bluff(2,2) is a multi-round game, the expected values of Bluff(1,1) are precomputed for payoffs at the leaves of Bluff(2,1), which is then solved for leaf payoffs in the full Bluff(2,2) game.  
In Bluff(2,2), the total number of histories $|H| \approx 10^{10}$. 

\subsection{Results}

\begin{table}
\begin{center}
\caption{PTTT and Bluff game sizes and properties. 
\label{tbl:sizes}}
\vspace{0.1in}
\begin{tabular}{|r|r|r|r|r|}
\hline
{\bf Game} & {\bf Abstr.} & {\bf Well-for.} & {\bf $|\mathcal{A}|$} & {\bf Savings}\\
\hline
DRP   & None     & Yes & 2610 & --- \\
DRP   & DRP-IR & Yes & 860 & 67.05\%  \\
\hline
PTTT  & None     & Yes &  11695314 & --- \\
PTTT  & FOSF      & Yes &   9347010 & 20.08\% \\
PTTT  & FOI       & No  &   1147530 & 90.19\% \\
PTTT  & FOS     & No  &   1484168 & 87.31\% \\
PTTT  & FOE      & No  &     47818 & 99.59\% \\
\hline
Bluff & None     & Yes & 704643030 & --- \\
Bluff & $r = 10$ & No  & 295534218 & 58.06\% \\
Bluff & $r =  8$ & No  & 108323418 & 84.63\% \\
Bluff & $r =  6$ & No  &  22518468 & 96.80\% \\
Bluff & $r =  4$ & No  &   2329068 & 99.67\% \\
Bluff & $r =  3$ & No  &    543900 & 99.92\% \\
Bluff & $r =  2$ & No  &     97608 & 99.97\% \\
Bluff & $r =  1$ & No  &     12600 & 99.99\% \\
\hline
\end{tabular}
\end{center}
\vskip -0.1in
\end{table}

\iftechreport

\begin{figure}[t!]
  \centering
  \begin{tabular}{c}
  \includegraphics[width=0.62\textwidth]{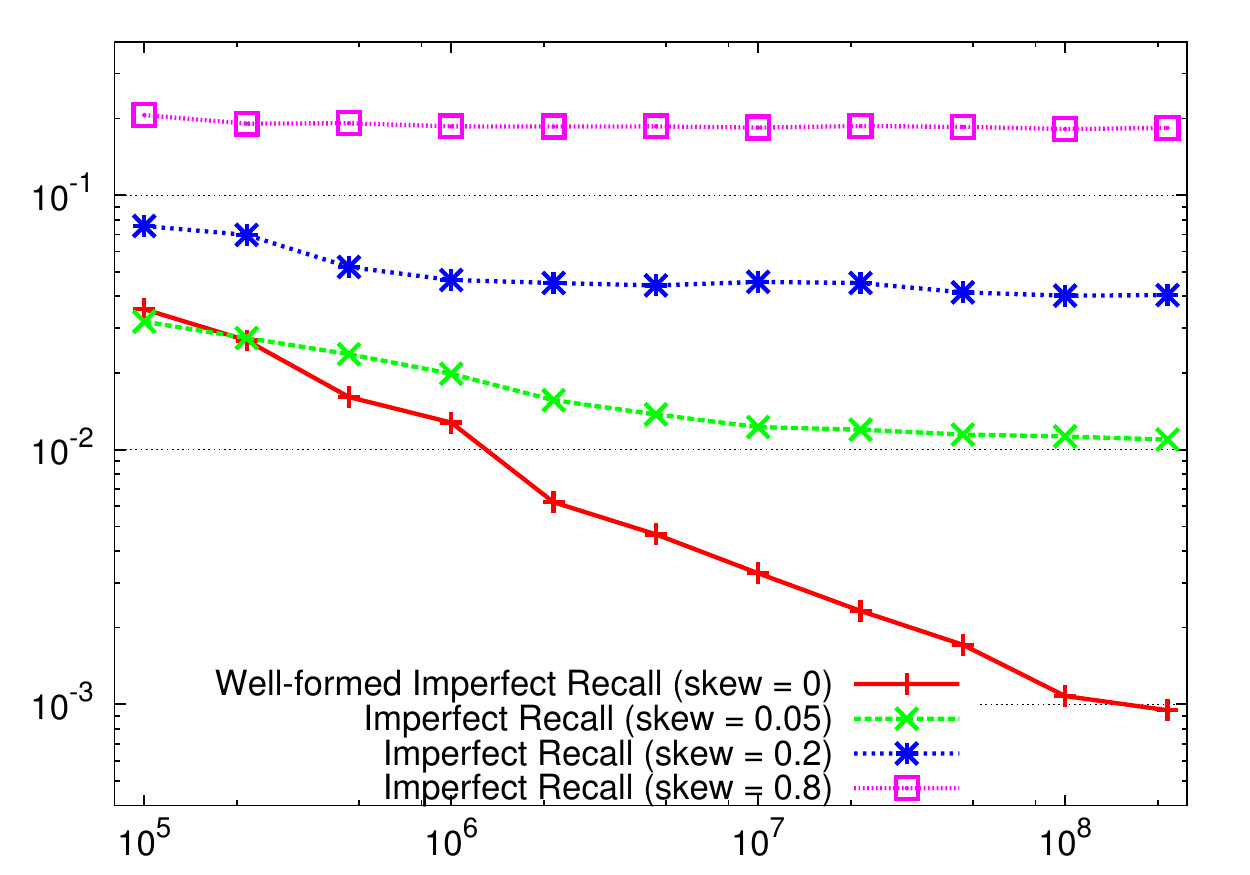} \\
  \includegraphics[width=0.62\textwidth]{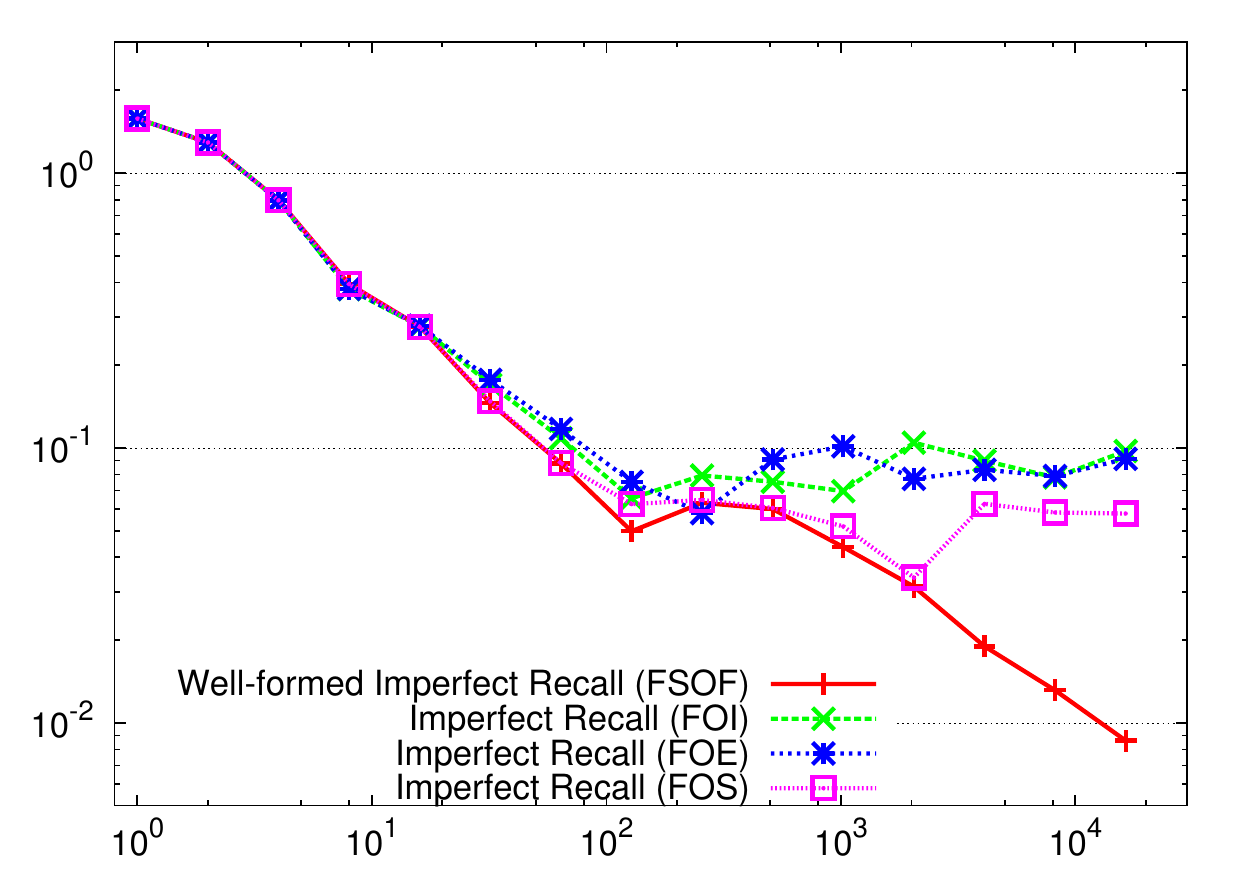} \\
  \includegraphics[width=0.62\textwidth]{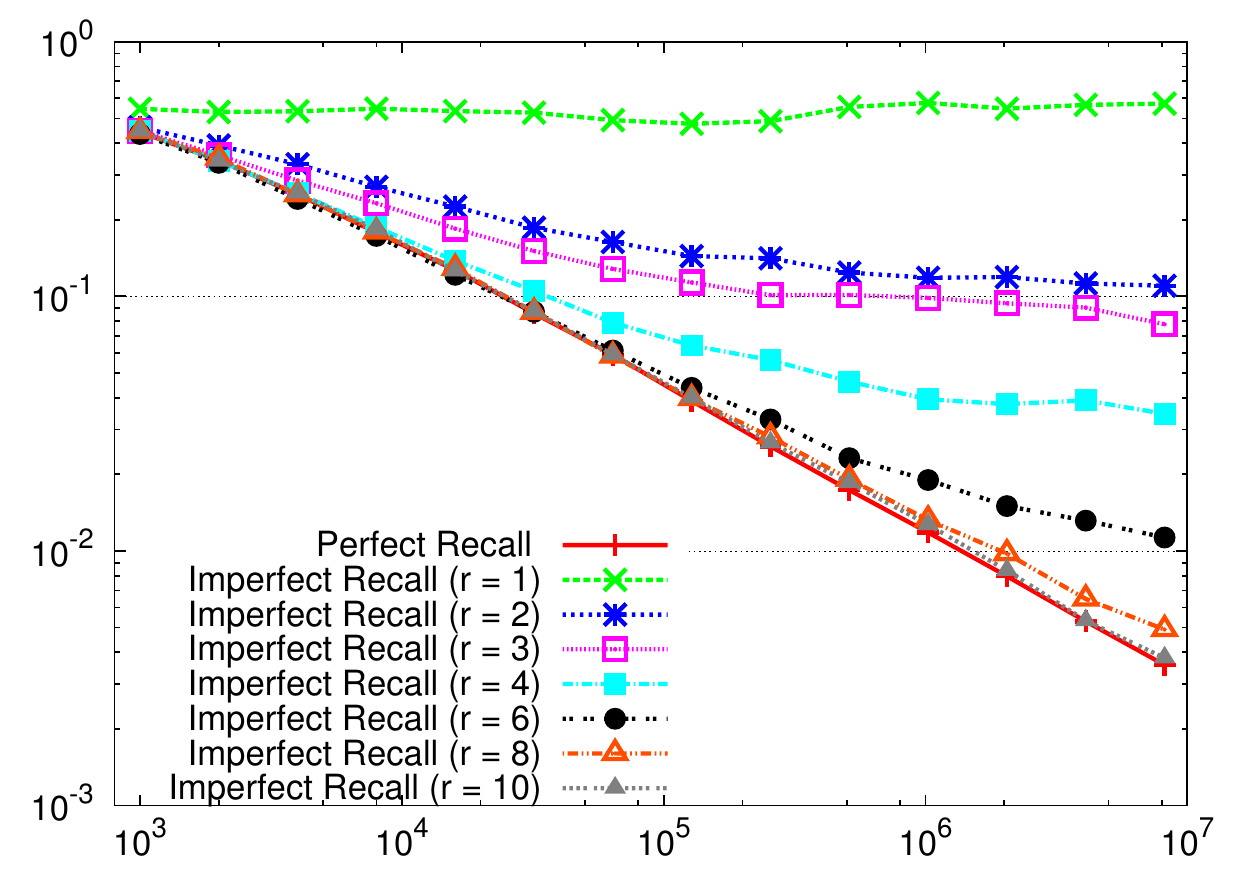} \\
  \end{tabular}
  \caption{Sum of average regrets for both players as iterations increase for Skew-DRP-IR($\delta$) (top),
             abstract games in PTTT (middle),
             and abstract games in Bluff (bottom). Each graph uses a log scale on both axes. 
             The vertical axes represent the sum of average regret for both players in the corresponding full, unabstracted game, and horizontal axes represent iterations.}
  \label{fig:convrates}
\end{figure}

\else

\begin{figure}[t!]
  \centering
  \begin{tabular}{c}
  \includegraphics[width=0.45\textwidth]{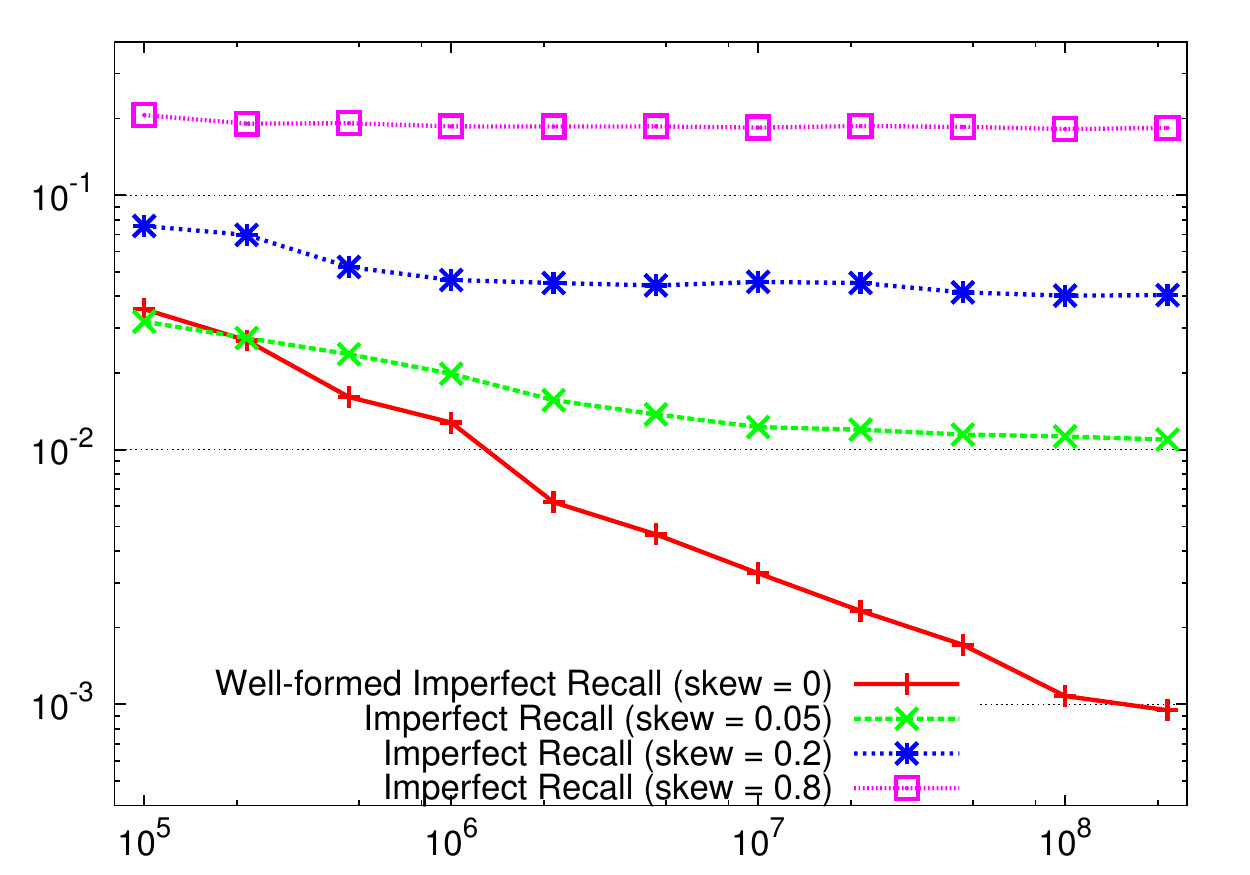} \\
  \includegraphics[width=0.45\textwidth]{figures/pttt} \\
  \includegraphics[width=0.45\textwidth]{figures/bluff} \\
  \end{tabular}
  \caption{Sum of average regrets for both players as iterations increase for Skew-DRP-IR($\delta$) (top),
             abstract games in PTTT (middle),
             and abstract games in Bluff (bottom). Each graph uses a log scale on both axes. 
             The vertical axes represent the sum of average regret for both players in the corresponding full, unabstracted game, and horizontal axes represent iterations.}
  \label{fig:convrates}
\end{figure}

\fi

We consider several different imperfect recall abstractions for DRP, Skew-DRP($\delta$), PTTT, and Bluff. 
For the DRP games, we apply DRP-IR and Skew-DRP-IR($\delta$) respectively as described in Section \ref{sec:theory}. 
Our PTTT and Bluff experiments, however, also investigate the effects of imperfect recall beyond skew well-formed games. 
In the full, perfect recall version of PTTT, each player remembers the order of every failed and every successful move she makes throughout the entire game. 
In our first abstract game, \defword{FOSF}, players forget the order of successive failures within the same turn. 
Clearly, there is an isomorphism between any two merged information sets $\breve{I}, \breve{I}' \in \breve{\mathcal{P}}(I)$ since the order of the actions does not affect the available future moves or utilities. 
Players still remember which turn each success and each failure occurred, and so the opponent's sequences of actions must be equal across the isomorphism. 
Thus, FOSF is well-formed. 
Our remaining PTTT abstractions, however, are not even skew well-formed. 
In \defword{FOI}, players independently remember the sequence of failures and the sequence of successful actions, but not how the actions interleave. 
In \defword{FOS}, players remember the order of failed actions, but not the order of successes. 
Finally, in \defword{FOE}, players only know what actions they have taken and remember nothing about the order in which they were taken. FOI, FOS, and FOE are not skew well-formed because no isomorphism can preserve the order of the opponent's previous information set, action pairs (breaking condition (iii) of Definitions \ref{def:wellformed} and \ref{def:skewwellformed}). 
In Bluff, we use abstractions described by Neller and Hnath~(\citeyear{Neller11}) that force players to forget everything except the last $r$ bids. 
Similarly, these abstract games are not skew well-formed because the players forget information that the opponent could previously distinguish. 
The size of each DRP, PTTT, and Bluff game is given in Table~\ref{tbl:sizes}.  
Here, $\mathcal{A} = \{ (I,a) : i \in N, I \in \II_i, a \in A(I) \}$ is the set of all information set, action pairs. 
Note that Skew-DRP($\delta$) is the same size as DRP regardless of the skew, and recall that CFR requires space linear in $\abs{\mathcal{A}}$. 


For each game, we ran CFR\footnote{Similar to Zinkevich \etal~(\citeyear{CFR}), we used the chance sampling variant of CFR.} on both players, meaning that each player's opponent was an identical copy of the same no-regret learner. 
The sum of the average regrets for each player over number of iterations is shown in Figure~\ref{fig:convrates}. 
The Skew-DRP-IR($\delta$) experiments show that as $\delta$ increases, so does the regret as predicted by Theorem \ref{thm:skew}, though $\sum_{I \in \mathcal{I}_i} \abs{\breve{\mathcal{P}}(I)} \delta_I$ appears to be a very loose bound on the final regret. 
In PTTT, regret diverges from zero for FOI, FOS, and FOE, where FOS appears to provide slightly better strategies than FOI and FOE.
While our theory cannot explain why FOS performs better, this does match our intuition that remembering information about the opponent's moves is important.  
For a small increase in average regret, FOS reduces the space required by 87\% compared to FOSF's 20\% reduction. 
Note that for both DRP and PTTT, running CFR on the full, perfect recall game achieves the same regret as in the well-formed abstractions (Skew-DRP-IR(0) and FSOF) and is thus not shown. 
In Bluff, we see that regret consistently worsens as fewer previous bids are remembered. 
This suggests that a result similar to Theorem \ref{thm:skew} for skew-well-formed games may hold if condition (iii) of Definition \ref{def:wellformed} is less constrained, though the proper formulation for such a relaxation remains unclear. 
Nonetheless, choosing $r=8$ saves 85\% of the memory with only a very small increase in average regret after millions of iterations. 

\section{Discussion}
\label{sec:discussion}

Well-formed games are described by four conditions provided in Definition \ref{def:wellformed}. 
Recall that Koller \& Megiddo~(\citeyear{Koller92the}) prove that determining a player's guaranteed payoff in an imperfect recall game is NP-complete. 
However, Koller \& Megiddo's NP-hardness reduction creates an imperfect recall game that breaks conditions (i), (iii), and (iv) of Definition \ref{def:wellformed}. 
In this section, we discuss the following question: For minimizing regret, how important is it to satisfy each individual condition of Definition \ref{def:wellformed}? 

Skew well-formed games and Theorem \ref{thm:skew} show that one can relax condition (i) of Definition \ref{def:wellformed} and still derive a bound on the average regret. 
In addition, most of our PTTT and Bluff abstractions from the previous section do not satisfy condition (iii), but CFR still produces reliable results. 
This suggests that it may be possible to relax condition (iii) in a similar manner to the relaxation of condition (i) introduced by skew well-formed games. 
While we leave this question open, we now demonstrate that breaking condition (iii) can lead CFR to a dead-lock situation where one player has constant average regret. 

\iftechreport

\begin{figure}[t!]
	\centering
	\includegraphics[width=0.75\textwidth,clip]{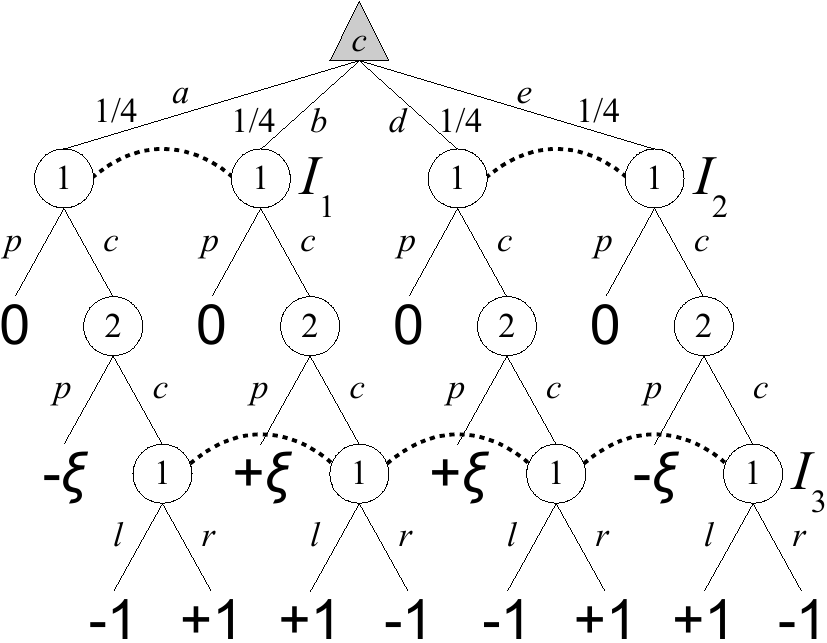}
	\caption{A zero-sum game with imperfect recall where CFR does not minimize average regret.  The utilities for player 1 are given at the terminal histories, where $\xi \in (0,1)$.  Nodes connected by a bold, dashed curve are in the same information set for player 1 (player 2 has perfect information).}
	\label{fig:cntrex}
\end{figure}

\else 

\begin{figure}[t!]
	\centering
	\includegraphics[width=0.48\textwidth,clip]{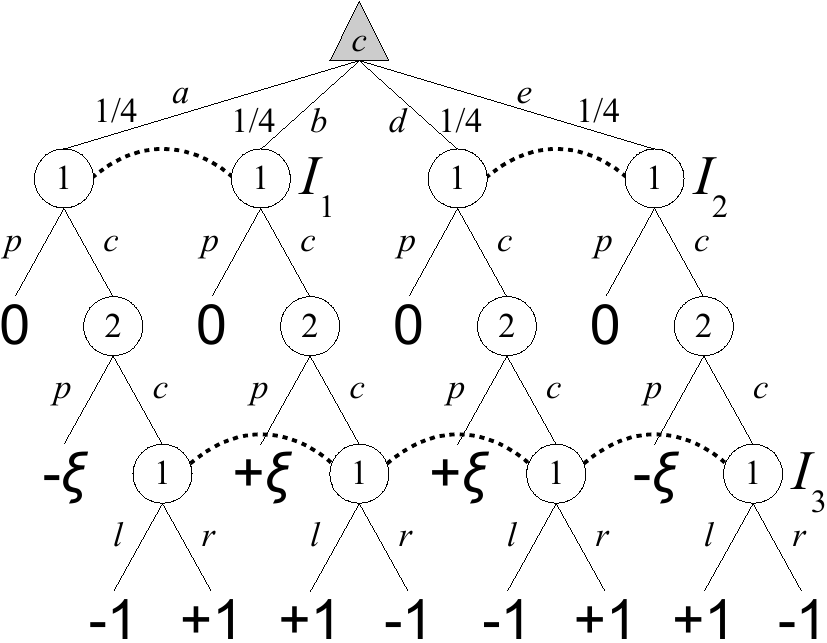}
	\caption{A zero-sum game with imperfect recall where CFR does not minimize average regret.  The utilities for player 1 are given at the terminal histories, where $\xi \in (0,1)$.  Nodes connected by a bold, dashed curve are in the same information set for player 1 (player 2 has perfect information).}
	\label{fig:cntrex}
\end{figure}

\fi

Let us walk through the process of applying CFR to the game in Figure~\ref{fig:cntrex}. 
Note that this game satisfies all of the conditions of Definition \ref{def:wellformed}, except for condition (iii). 
To begin, the current strategy profile $\sigma^1$ is set to be uniform random at every information set.
Under this profile, when player 1 is at $I_3$, each of the four histories are equally likely.
Thus, $v_i(\sigma_{(I_3 \rightarrow l)}^1, I_3) = v_i(\sigma_{(I_3 \rightarrow r)}^1, I_3) = v_i(\sigma^1, I_3) = 0$, and so $r_1^1(I_3, l) = r_1^1(I_3, r) = 0$.
In addition, under $\sigma^1$, the counterfactual value of the pass ($p$) and continue ($c$) actions at both $I_1$ and $I_2$ is zero, and thus the immediate counterfactual regrets at $I_1$ and $I_2$ on iteration 1 are also zero.
Player 2, however, has positive immediate counterfactual regret for passing ($p$) at histories $ac$ and $ec$ (to always receive $\xi$ utility) and for continuing ($c$) at $bc$ and $de$ (to always avoid receiving $-\xi$ utility), and has negative immediate counterfactual regret for continuing at $ac$ and $ec$ and for passing at $bc$ and $de$.
Therefore, the next profile $\sigma^2$ still has player 1 playing uniformly random everywhere, but player 2 now always passes at $ac$ and $ec$, and always continues at $bc$ and $dc$.
On the second iteration of CFR, the positive regrets for player 1 at $I_3$ remain the same because the histories $bcc$ and $dcc$ are equally likely.  
Also, player 2's positive regrets remain the same at all four histories in $H_2$.
However, player 1's expected utility for continuing at $I_1$ or $I_2$ is now negative since player 2 now passes at $ac$ and $ec$.
Thus, player 1 gains positive regret for passing at both $I_1$ and $I_2$.
This leads us to the next profile $\sigma^3 = \{ (I_1,p) = 1, (I_2,p) = 1, (ac, p) = 1, (bc, p) = 0, (dc, p) = 0, (ec, p) = 1, (I_3,l) = 0.5 \}$.
One can check that running CFR for more iterations yields $\sigma^t = \sigma^3$ for all $t \geq 3$. 
The average regret for playing this way will be constant and hence does not approach zero because  
player 1 would rather play $\sigma_1' = \{ (I_1, p) = 1, (I_2, p) = 0, (I_3, l) = 0 \}$ and get 
$u_1(\sigma_1', \sigma_2^3) = (1 - \xi) / 4 > u_1(\sigma^3)$ for $\xi \in (0,1)$.
A similar example can be constructed where condition (iii) holds, but chance's probabilities are not proportional (breaking condition (ii)). 

Despite the problem of breaking condition (iii), condition (iv) of Definition~\ref{def:wellformed} can be relaxed. 
Rather than enforcing player $i$'s future information to be the same across the bijection $\phi$, we only require that the corresponding subtrees be isomorphic, allowing player $i$ to re-remember information that was previously forgotten. 
The details for this relaxation are in the supplementary material. 
However, it is not clear that this relaxation is possible in skew well-formed games, nor does it seem to provide any practical advantage. 

\section{Conclusion}

We have provided the first set of theoretical guarantees for CFR in imperfect recall games. 
We defined well-formed and skew well-formed games and provided bounds on the average regret that results from applying CFR to such games. 
In addition, our theory shows that we can achieve low average regret in a full, perfect recall game when employing CFR on an abstract version of the game, provided the abstract game is skew well-formed (with or without imperfect recall). 
Our DRP experiments confirm these theoretical results, while our PTTT and Bluff experiments hint that it may be possible to still bound regret in other types of imperfect recall games. 
Future work will look to expand on the set of imperfect recall games to which CFR can be reliably applied. 
In particular, it may be possible to derive regret bounds for a new class of games where conditions (ii) and (iii) of Definition \ref{def:wellformed} are relaxed.

\section*{Acknowledgments}

We would like to thank the Computer Poker Research Group at the University of Alberta for their helpful discussions that contributed to this work. 
This work was supported by NSERC, Alberta Innovates -- Technology Futures, and the use of computing resources provided by WestGrid and Compute Canada. 

\nocite{MCCFR-TR}

\bibliography{paper}
\bibliographystyle{plainnat}

\appendix

\section*{Appendix A}

In this section, we will prove Theorems 1 and 2 of the main paper.  
Note that by the definition of counterfactual value, the regrets between $\Gamma$ and a perfect recall refinement $\breve{\Gamma}$ are additive; specifically, for $I \in \mathcal{I}_i$ in $\Gamma$,
\begin{equation}
\label{eq:addReg}
R_i^T(I,a) = \sum_{\breve{I} \in \breve{\mathcal{P}}(I)} R_i^T(\breve{I}, a).
\end{equation}

First, we provide a lemma that generalizes Theorem 4 of (Zinkevich et al., 2008) by showing that if the immediate counterfactual regrets of each $\breve{I} \in \breve{\mathcal{P}}(I)$ are proportional up to some difference $D$, then the average regret can be bounded above:
\begin{lemmaA}
Let $\breve{\Gamma}$ be a perfect recall refinement of a game $\Gamma$.  If for all $I \in \mathcal{I}_i$, $\breve{I}, \breve{I}' \in \breve{\mathcal{P}}(I)$, and $a \in A(I)$, there exist constants $C_{\breve{I},\breve{I}',a}, D_{\breve{I},\breve{I}',a} \in [0, \infty)$ such that
\begin{equation}
\label{eq:lemUnsafeCond}
\frac{1}{T}\left| R_i^{T,+}(\breve{I}, a) - C_{\breve{I},\breve{I}',a}R_i^{T,+}(\breve{I}', a) \right| \leq D_{\breve{I},\breve{I}',a},
\end{equation}
then the average regret in $\breve{\Gamma}$ is bounded by
\[ \frac{\breve{R}_i^T}{T} \leq \frac{\Delta_i C \sqrt{|A_i|} }{\sqrt{T}} + \sum_{I \in \mathcal{I}} |\breve{\mathcal{P}}(I)|D_I, \]
where 
\[ C = \sum_{I \in \mathcal{I}_i} \max_{\breve{I},\breve{I}' \in \breve{\mathcal{P}}(I), a \in A(I)} C_{\breve{I},\breve{I}',a}\] 
and 
\[ D_I = \max_{\breve{I},\breve{I}' \in \breve{\mathcal{P}}(I), a \in A(I)} D_{\breve{I},\breve{I}',a}. \]
\end{lemmaA}
\begin{Proof}
\begin{align*}
\breve{R}_i^T &\leq \sum_{\breve{I} \in \breve{\mathcal{I}}_i} \max_{a \in A(I)} R_i^{T,+}(\breve{I}, a) \text{ by Theorem 3 of (Zinkevich et al., 2008)} \\
&= \sum_{I \in \mathcal{I}_i} \sum_{\breve{I} \in \breve{\mathcal{P}}(I)} \max_{a \in A(I)} R_i^{T,+}(\breve{I}, a) \text{ by definition of a perfect recall refinement} \\
&\leq \sum_{I \in \mathcal{I}_i} |\breve{\mathcal{P}}(I)| R_i^{T,+}(\breve{I}^*, a^*) \text{ where } \breve{I}^* = \argmax_{\breve{I} \in \breve{\mathcal{P}}(I)} \max_{a \in A(I)} R_i^{T,+}(\breve{I}, a) \\
&\ \ \ \ \ \text{and } a^* = \argmax_{a \in A(I)} R_i^{T,+}(\breve{I}^*, a) \\
&\leq \sum_{I \in \mathcal{I}_i} |\breve{\mathcal{P}}(I)| \left( C_{\breve{I}^*,\breve{I}^{**},a^*} R_i^{T,+}(\breve{I}^{**}, a^*) + TD_{\breve{I}^*,\breve{I}^{**},a^*} \right) \text{ by \eqref{eq:lemUnsafeCond},} \\
& \ \ \ \ \ \text{where } \breve{I}^{**} = \argmin_{\breve{I} \in \breve{\mathcal{P}}(I)} R_i^T(\breve{I}, a^*) \\
&\leq \sum_{I \in \mathcal{I}_i} |\breve{\mathcal{P}}(I)| C_{\breve{I}^*,\breve{I}^{**},a^*} \left( \frac{1}{|\breve{\mathcal{P}}(I)|} \sum_{\breve{I} \in \breve{\mathcal{P}}(I)} R_i^{T}(\breve{I}, a^*) \right)^+ + T\sum_{I \in \mathcal{I}_i} |\breve{\mathcal{P}}(I)| D_I \\
&\ \ \ \ \ \text{because the minimum is less than the average and $(\cdot)^+$ is monotone increasing} \\
&= \sum_{I \in \mathcal{I}_i} C_{\breve{I}^*,\breve{I}^{**},a^*} R_i^{T,+}(I, a^*) + T\sum_{I \in \mathcal{I}_i} |\breve{\mathcal{P}}(I)| D_I \text{ by \eqref{eq:addReg}} \\
&\leq \sum_{I \in \mathcal{I}_i} C_{\breve{I}^*,\breve{I}^{**},a^*} T\sqrt{\sum_{a \in A(I)} \left( \frac{R_i^{T,+}(I,a)}{T} \right) ^2 } + T\sum_{I \in \mathcal{I}_i} |\breve{\mathcal{P}}(I)| D_I \\
&\leq \sum_{I \in \mathcal{I}_i} C_{\breve{I}^*,\breve{I}^{**},a^*} \Delta_i \sqrt{|A(I)|}\sqrt{T} + T\sum_{I \in \mathcal{I}_i} |\breve{\mathcal{P}}(I)| D_I \\
&\ \ \ \ \ \text{by Theorem 6 of (Lanctot et al., 2009)} \\
&\leq \Delta_i C\sqrt{|A_i|}\sqrt{T} + T\sum_{I \in \mathcal{I}_i} |\breve{\mathcal{P}}(I)| D_I.
\end{align*}
Dividing both sides by $T$ establishes the lemma. \Qed
\end{Proof}


\noindent Note that if $\Gamma$ has perfect recall, then the constants $C_{I,I,a} = 1$ and $D_{I,I,a} = 0$ for all $I \in \mathcal{I}_i$ and $a \in A(I)$ satisfy the condition of Lemma A.
In this case, $C = |\mathcal{I}_i|$ and $D_I = 0$, and so $R_i^T / T \leq \Delta_i |\mathcal{I}_i| \sqrt{|A_i|} / \sqrt{T}$, recovering Theorem 4 of (Zinkevich et al., 2008).

We now use Lemma A to prove Theorems 1 and 2:
\begin{theorem2}
\label{thm:skew}
If $\Gamma$ is skew well-formed with respect to $\breve{\Gamma}$, then the average regret in $\breve{\Gamma}$ for player $i$ of choosing strategies according to CFR in $\Gamma$ is bounded by
\[ \frac{\breve{R}_i^T}{T} \leq \frac{\Delta_{i} K \sqrt{|A_i|}}{\sqrt{T}} + \sum_{I \in \mathcal{I}_i} |\breve{\mathcal{P}}(I)| \delta_I, \]
where $K = \sum_{I \in \mathcal{I}_i} \max_{\breve{I},\breve{I}' \in \breve{\mathcal{P}}(I)} k_{\breve{I},\breve{I}'} \ell_{\breve{I},\breve{I}'}$ and $\delta_I = \max_{\breve{I},\breve{I}' \in \breve{\mathcal{P}}(I)} \delta_{\breve{I},\breve{I}'}\ell_{\breve{I},\breve{I}'}$.
\end{theorem2}
\begin{Proof}
We will show that for all $I \in \mathcal{I}_i$, $\breve{I}, \breve{I}' \in \breve{\mathcal{P}}(I)$, and $a \in A(I)$,
\begin{equation}
\label{eq:unsafegoal}
\frac{1}{T} \left| R_i^{T,+}(\breve{I}, a) - k_{\breve{I},\breve{I}'}\ell_{\breve{I},\breve{I}'}R_i^{T,+}(\breve{I}', a) \right| \leq \delta_{\breve{I},\breve{I}'}\ell_{\breve{I},\breve{I}'},
\end{equation}
which, by Lemma A, proves the theorem.

Fix $I \in \mathcal{I}_i$, $\breve{I}, \breve{I}' \in \breve{\mathcal{P}}(I)$, and $a \in A(I)$.  
Firstly, for all $z \in Z_{\breve{I}}$ and $\sigma \in \Sigma$, by conditions (ii) and (iii) of Definition 3, we have 
\begin{align}
\pi_{-i}^\sigma(z) &= \pi_c(z) \prod_{(I,a) \in X_{-i}(z)} \sigma(I,a) \nonumber \\
&= \ell_{\breve{I},\breve{I}'} \pi_c(\phi(z)) \prod_{(I,a) \in X_{-i}(\phi(z))} \sigma(I,a) \nonumber \\
&= \ell_{\breve{I},\breve{I}'}\pi_{-i}^\sigma(\phi(z)) \label{eq:unsafeproof1}
\end{align}
and by condition (iv) of Definition 3, we similarly have
\begin{equation}
\label{eq:unsafeproof2}
\pi_{i}^\sigma(z[\breve{I}], z) = \pi_{i}^\sigma(\phi(z)[\breve{I}'], \phi(z)) 
\end{equation}
and
\begin{equation}
\label{eq:unsafeproof3}
\pi_{i}^\sigma(z[\breve{I}]a, z) = \pi_{i}^\sigma(\phi(z)[\breve{I}']a, \phi(z)).
\end{equation}

We can then bound the positive part of the immediate counterfactual regret $R_i^{T,+}(\breve{I},a)$ above by

\begin{align*}
R_i^{T,+}(\breve{I}, a) &= \left( \sum_{t=1}^T r_i^t(\breve{I},a) \right)^+ \nonumber \\
&= \left( \sum_{t=1}^T \sum_{z \in Z_{\breve{I}}} \pi_{-i}^\sigma(z) (\pi_i^{\sigma}(z[\breve{I}]a,z) - \pi_i^{\sigma}(z[\breve{I}],z) ) u_i(z) \right)^+ \nonumber \\
&\leq \mbox{\Huge $($} \sum_{t=1}^T \sum_{z \in Z_{\breve{I}}} \ell_{\breve{I},\breve{I}'} \pi_{-i}^\sigma(\phi(z)) (\pi_i^{\sigma}(\phi(z)[\breve{I}']a,\phi(z)) \nonumber \\ 
&\ \ \ \ \ \ \ \ \ \ \ \ \ \ \ \ \ \ \ \ \ \ \ \ \ - \pi_i^{\sigma}(\phi(z)[\breve{I}'],\phi(z)) ) (k_{\breve{I},\breve{I}'}u_i(\phi(z)) + \delta_{\breve{I},\breve{I}'} )\mbox{\Huge $)$}^+ \nonumber \\
&\ \ \ \ \ \text{by equations \eqref{eq:unsafeproof1}, \eqref{eq:unsafeproof2}, \eqref{eq:unsafeproof3}, and condition (i) of Definition 3} \nonumber \\
\end{align*}

\begin{align}
&=  \mbox{\Huge $($} \sum_{t=1}^T \sum_{z \in Z_{\breve{I}'}} \ell_{\breve{I},\breve{I}'} \pi_{-i}^\sigma(z) (\pi_i^{\sigma}(z[\breve{I}']a,z) \nonumber \\
&\ \ \ \ \ \ \ \ \ \ \ \ \ \ \ \ \ \ \ \ \ \ \ \ \ - \pi_i^{\sigma}(z[\breve{I}'],z) ) (k_{\breve{I},\breve{I}'}u_i(z) + \delta_{\breve{I},\breve{I}'} ) \mbox{\Huge $)$}^+ \nonumber \\
&\ \ \ \ \ \text{since $\phi$ is a bijection} \nonumber \\
&\leq \left( \sum_{t=1}^T \sum_{z \in Z_{\breve{I}'}} k_{\breve{I},\breve{I}'}\ell_{\breve{I},\breve{I}'} \pi_{-i}^\sigma(z) (\pi_i^{\sigma}(z[\breve{I}]a,z) - \pi_i^{\sigma}(z[\breve{I}],z) ) u_i(z) \right)^+ \nonumber \\
&\ \ \ \ \ + \left( \sum_{t=1}^T \sum_{z \in Z_{\breve{I}'}} \delta_{\breve{I},\breve{I}'}\ell_{\breve{I},\breve{I}'} \pi_{-i}^\sigma(z) (\pi_i^{\sigma}(z[\breve{I}]a,z) - \pi_i^{\sigma}(z[\breve{I}],z) )\right)^+ \nonumber \\
&\leq k_{\breve{I},\breve{I}'}\ell_{\breve{I},\breve{I}'} R_i^{T,+}(\breve{I}', a) + \sum_{t=1}^T \delta_{\breve{I},\breve{I}'}\ell_{\breve{I},\breve{I}'} \pi_{-i}^\sigma(\breve{I}') \nonumber \\
&\leq k_{\breve{I},\breve{I}'}\ell_{\breve{I},\breve{I}'} R_i^{T,+}(\breve{I}', a) + T\delta_{\breve{I},\breve{I}'}\ell_{\breve{I},\breve{I}'} \label{eq:unsafeproof4},
\end{align}

where the last line follows because $\pi_{-i}^\sigma(\breve{I}') = \sum_{z \in Z_{\breve{I}'}} \pi_{-i}^\sigma(z[\breve{I}']) \leq 1$ in a perfect recall game $\breve{\Gamma}$.
Similarly,

\begin{equation}
\label{eq:unsafeproof5}
R_i^{T,+}(\breve{I}, a) \geq k_{\breve{I},\breve{I}'}\ell_{\breve{I},\breve{I}'} R_i^{T,+}(\breve{I}', a) - T\delta_{\breve{I},\breve{I}'}\ell_{\breve{I},\breve{I}'}, 
\end{equation}
which together with equation \eqref{eq:unsafeproof4} and dividing by $T$ establishes \eqref{eq:unsafegoal}, completing the proof. \Qed
\end{Proof}

\noindent Note that Theorem 1 immediately follows from Theorem 2 since a well-formed game is skew well-formed with $\delta_{\breve{I},\breve{I}'} = 0$ for all $\breve{I}, \breve{I}' \in \breve{\mathcal{P}}(I)$. 

\section*{Appendix B}

In this section, we consider an alternative extension of well-formed games that relaxes condition (iv) of Definition 2. 
For a subset of histories $S \subseteq H_i$, define 
\[ D_i(S) = \{I \mid I \in \mathcal{I}_i, \exists h \in S, h' \in I \text{ such that } h \sqsubseteq h'\}\] 
to be the set of all information sets descending from any history in $S$.
\begin{definition4}
For a game $\Gamma$ and a perfect recall refinement $\check{\Gamma}$, we say that $\Gamma$ is a \defword{nearly well-formed game with respect to $\check{\Gamma}$} if for all $i \in N$, $I \in \mathcal{I}_i$, $\check{I}, \check{I}' \in \check{\mathcal{P}}(I), J \in D_i(\check{I})$, there exist bijections $\phi: Z_{\check{I}} \rightarrow Z_{\check{I}'}$, $\psi: D_i(\check{I}) \rightarrow D_i(\check{I}')$, $\omega: A(J) \rightarrow A(\psi(J))$ and constants $k_{\check{I},\check{I}'}, \ell_{\check{I},\check{I}'} \in [0, \infty)$ such that for all $z \in Z_{\check{I}}$:
\begin{enumerate}
\item[\emph{(i)}] $u_i(z) = k_{\check{I},\check{I}'}u_i(\phi(z))$,
\item[\emph{(ii)}] $\pi_c(z) = \ell_{\check{I},\check{I}'} \pi_c(\phi(z))$,
\item[\emph{(iii)}] In $\Gamma$, $X_{-i}(z) = X_{-i}(\phi(z))$, and 
\item[\emph{(iv)}] $X_i(z[\check{I}], z) = (J_1, a_1), ..., (J_m, a_m)$ if and only if \\$X_i(\phi(z)[\check{I}'], \phi(z))= (\psi(J_1), \omega(a_1)), ..., (\psi(J_m), \omega(a_m))$.
\end{enumerate}
We say that $\Gamma$ is a \defword{nearly well-formed game} if it is nearly well-formed with respect to some perfect recall refinement.
\end{definition4}
In a nearly well-formed game, condition (iv) says that player $i$ may now remember information that was once forgotten, provided the descendants from $\check{I}$ and $\check{I}'$ are isomorphic across $\phi$. 
This relaxes the corresponding condition for a well-formed game where player $i$ could never remember information once it was forgotten. 
Clearly, any well-formed game is nearly well-formed by choosing $\psi$ and $\omega$ to be the identity bijections.

For example, consider a longer version of DRP, \defword{DRP-3}, that consists of three betting rounds instead of two where a third die is rolled at the beginning of round 3. 
We then define \defword{DRP-IR-3} to be the imperfect recall abstraction of DRP-3 where during round 2, players only know the sum of their two dice. 
In round 3, players once again know the outcome of each individual die roll, recovering information from the first round that was forgotten in the second. 
For instance, corresponding histories where player $i$'s first two rolls were 1,5 and where her first two rolls were 4,2 will be in the same information set during round 2, but will be in different information sets in round 3. 
However, betting is independent of dice rolls and utilities are only dependent on the final sum of the three dice. 
Therefore, the descendants from these histories are isomorphic across $\phi$ and thus DRP-IR-3 is nearly well-formed with respect to DRP-3.

CFR guarantees that the average regret is also minimized in nearly well-formed games:
\begin{theorem3}
If $\Gamma$ is nearly well-formed with respect to $\breve{\Gamma}$, then the average regret in $\breve{\Gamma}$ for player $i$ of choosing strategies according to CFR in $\Gamma$ is bounded by
\[ \frac{\check{R}_i^T}{T} \leq \frac{\Delta_i K \sqrt{|A_i|}}{\sqrt{T}}, \]
where $K = \sum_{I \in \mathcal{I}_i} \max_{\check{I},\check{I}' \in \check{\mathcal{P}}(I)} k_{\check{I},\check{I}'} \ell_{\check{I},\check{I}'}$.
\end{theorem3}
\begin{Proof}
Fix $I \in \mathcal{I}_i$, $\breve{I}, \breve{I}' \in \breve{\mathcal{P}}(I)$, and $a \in A(I)$. 
By conditions (ii) and (iii) of Definition 4, equation \eqref{eq:unsafeproof1} holds.

\bigskip
\noindent \textbf{Claim:} $R_i^T(J,b) = k_{\breve{I},\breve{I}'}\ell_{\breve{I},\breve{I}'}R_i^T(\psi(J), \omega(b))$ for all $J \in D_i(\breve{I})$, $b \in A(J)$, $T \geq 0$.
\bigskip

\noindent Provided the claim is true, we have 
\begin{align}
\sigma^{T+1}(J,b) &= \left\{ \begin{array}{ll} \frac{R_i^{T,+}(J,b)}{\sum_{d \in A(J)} R_i^{T,+}(J,d)} & \text{if } \sum_{d \in A(J)} R_i^{T,+}(J,d) > 0 \\ \frac{1}{|A(J)|} & \text{otherwise} \end{array} \right. \nonumber \\
&= \left\{ \begin{array}{l} \frac{k_{\breve{I},\breve{I}'}\ell_{\breve{I},\breve{I}'}R_i^{T,+}(\psi(J),\omega(b))}{\sum_{d \in A(J)} k_{\breve{I},\breve{I}'}\ell_{\breve{I},\breve{I}'}R_i^{T,+}(\psi(J),\omega(b))} \\\ \ \ \ \ \ \ \ \ \ \ \text{if } \sum_{d \in A(J)} k_{\breve{I},\breve{I}'}\ell_{\breve{I},\breve{I}'}R_i^{T,+}(\psi(J),\omega(b)) > 0 \nonumber \\ \frac{1}{|A(\psi(J))|} \ \ \ \ \ \text{ otherwise} \end{array} \right. \nonumber \\
&\ \ \ \ \ \text{since $\omega$ is a bijection} \nonumber \\
&= \sigma^{T+1}(\psi(J),\omega(b)) \label{eq:altproof1}
\end{align}
for all $J \in D_i(\breve{I})$, $b \in A(J)$, $T \geq 0$.
Therefore, for $t \geq 1$,
\begin{align*}
\pi_i^{\sigma^t}(z[\breve{I}], z) &= \prod_{(J,b) \in X_i(z[\breve{I}], z)} \sigma^t(J,b) \\
&= \prod_{(J,b) \in X_i(z[\breve{I}], z)} \sigma^t(\psi(J),\omega(b)) \\
&= \prod_{(J,b) \in X_i(\phi(z)[\breve{I}'], \phi(z))} \sigma^t(J, b) \text{ by condition (iv) of Definition 4 }\\
&= \pi_i^{\sigma^t}(\phi(z)[\breve{I}'], \phi(z)),
\end{align*}
and thus equation \eqref{eq:unsafeproof2} and similarly equation \eqref{eq:unsafeproof3} hold for $\sigma = \sigma^t$. 
By following the proof of Theorem 2, we then have that equations \eqref{eq:unsafeproof4} and \eqref{eq:unsafeproof5} with $\delta_{\breve{I}, \breve{I}'} = 0$ hold, and hence equation \eqref{eq:unsafegoal} with $\delta_{\breve{I}, \breve{I}'} = 0$ holds.  
This establishes the theorem by Lemma A.

To complete the proof, we are left to show that the claim holds.  
We will do so by induction on $T$.  
The base case $T=0$ holds since $R_i^0(I,a) = 0$ for all $I \in \mathcal{I}_i$, $a \in A(I)$.  
For the inductive step, assume that $R_i^{T-1}(J,b) = k_{\breve{I},\breve{I}'}\ell_{\breve{I},\breve{I}'}R_i^{T-1}(\psi(J), \omega(b))$ for all $J \in D_i(\breve{I})$, $b \in A(J)$. 
We will show that $R_i^{T}(J,b) = k_{\breve{I},\breve{I}'}\ell_{\breve{I},\breve{I}'}R_i^{T}(\psi(J), \omega(b))$ for all $J \in D_i(\breve{I})$, $b \in A(J)$.

Fix $J \in D_i(\breve{I})$ and $b \in A(J)$. 
By equation \eqref{eq:altproof1}, we have for all $z \in Z_J$,
\begin{align}
\pi_i^{\sigma^T}(z[J],z) &= \prod_{(J', b') \in X_i(z[J], z)} \sigma^T(J', b') \nonumber \\
&= \prod_{(J', b') \in X_i(z[J], z)} \sigma^T(\psi(J'), \omega(b')) \text{ by equation \eqref{eq:altproof1}} \nonumber \\
&= \prod_{(J', b') \in X_i(\phi(z)[\psi(J)], \phi(z))} \sigma^T(J', b') \nonumber \\
&\ \ \ \ \ \text{by condition (iv) of Definition 4 since $X_i(z[J], z)$ is a subsequence} \nonumber \\
&\ \ \ \ \ \text{(more precisely, a suffix) of $X_i(z[\breve{I}], z)$} \nonumber \\
&= \pi_i^{\sigma^T}(\phi(z)[\psi(J)], \phi(z)) \label{eq:altproof2}
\end{align}
and similarly 
\begin{equation}
\label{eq:altproof3}
\pi_i^{\sigma^T}(z[J]b,z) = \pi_i^{\sigma^T}(\phi(z)[\psi(J)]\omega(b), \phi(z)).
\end{equation}
Now consider the counterfactual regret at time $T$,
\begin{align*}
r_i^T(J,b) &= \sum_{z \in Z_J} \pi_{-i}^{\sigma^T}(z) ( \pi_i^{\sigma^T}(z[J]b, z) - \pi_i^{\sigma^T}(z[J],z)) u_i(z) \\
&= \sum_{z \in Z_J} \ell_{\breve{I},\breve{I}'} \pi_{-i}^{\sigma^T}(\phi(z)) ( \pi_i^{\sigma^T}(\phi(z)[\psi(J)]\omega(b), \phi(z)) \\
&\ \ \ \ \ \ \ \ \ \ \ \ \ \ \ - \pi_i^{\sigma^T}(\phi(z)[\psi(J)], \phi(z)) ) k_{\breve{I},\breve{I}'} u_i(\phi(z)) \\
&\ \ \ \ \ \text{by equations \eqref{eq:altproof2}, \eqref{eq:altproof3} and conditions (i), (ii), and (iii) of Definition 4} \\
&= \ell_{\breve{I},\breve{I}'}k_{\breve{I},\breve{I}'} r_i^T(\psi(J), \omega(b)).
\end{align*}
Finally,
\begin{align*}
R_i^T(J,b) &= \sum_{t=1}^T r_i^t(J,b) \\
&= R_i^{T-1}(J,b) + r_i^T(J,b) \\
&= \ell_{\breve{I},\breve{I}'}k_{\breve{I},\breve{I}'}( R_i^{T-1}(\psi(J), \omega(b)) + r_i^T(\psi(J), \omega(b)) ) \\
&\ \ \ \ \ \text{by the induction hypothesis and the above} \\
&= \ell_{\breve{I},\breve{I}'}k_{\breve{I},\breve{I}'} \sum_{t=1}^T r_i^t(\psi(J), \omega(b)) \\
&= \ell_{\breve{I},\breve{I}'}k_{\breve{I},\breve{I}'} R_i^T(\psi(J), \omega(b)),
\end{align*}
establishing the inductive step.  
This completes the proof. \Qed
\end{Proof}


\end{document}